\documentclass[aps,prd,amsmath,amssymb,nofootinbib,superscriptaddress,floatfix,10pt]{revtex4-2}
\usepackage{graphicx}
\usepackage{dcolumn}
\usepackage[utf8]{inputenc}
\usepackage{xcolor}
\usepackage{diagbox}
\usepackage[unicode=true,colorlinks,linkcolor=blue,anchorcolor=blue,urlcolor=blue,citecolor=blue,breaklinks=true]{hyperref}
\usepackage{subfigure}
\usepackage{epstopdf}
\usepackage{orcidlink}
\usepackage{multirow}
\usepackage{makecell}

\newcommand{\pararrow}{\mathord{\buildrel{\lower3pt\hbox{$\scriptscriptstyle\leftrightarrow$}}\over {\partial}}} % partial leftrightarrow operation
\newcommand{\pararrowk}[1]{\mathord{\buildrel{\lower3pt\hbox{$\scriptscriptstyle\leftrightarrow$}}\over {\partial}\hspace*{-0.18em}{}^#1}\hspace*{-0.18em} \,} % partial
\usepackage{bm}
\usepackage{slashed}

\newcommand{\zjwl}{\affiliation{College of Information and Intelligence Engineering, Zhejiang Wanli University, Zhejiang 315101, China}}

\newcommand{\nbu}{\affiliation{Physics Department, Ningbo University, Zhejiang 315211, China}}

\newcommand{\usc}{\affiliation{School of Nuclear Science and Technology, University of South China, Hengyang, 421001, Hunan, China}}

\newcommand{\kmu}{\affiliation{School of Physical Science and Technology, Kunming University, Kunming 650214, China}}

\begin{document}

\title{The spectra of $bc\bar{b}\bar{c}$ tetraquark states from a diquark-antidiquark perspective}
\author{Zhen-Yang Wang}
\email{wangzhenyang@nbu.edu.cn}
\nbu
\author{Jing-Juan Qi}
\email{qijj@mail.bnu.edu.cn}
\zjwl\nbu
\author{Zhen-Hua Zhang}
\email{zhangzh@usc.edu.cn}
\usc
\author{Xin-Heng Guo}
\email{corresponding author: xhguo@bnu.edu.cn}
\kmu

\begin{abstract}
Within the diquark-antidiquark framework, this study investigates the masses of ground-state tetraquarks composed of heavy charm ($c$) and bottom ($b$) quarks and antiquarks, using the Bethe-Salpeter formalism. We establish the Bethe-Salpeter equations for the fully heavy $bc\bar{b}\bar{c}$ tetraquarks to the leading order in $1/m_Q$ expansion. These equations are subsequently solved numerically under the covariant instantaneous approximation with kernels containing scalar confinement and one-gluon exchange terms. Our results show that the spectra of all possible $S$-wave tetraquark states are above the corresponding two lowest meson decay thresholds via the quark rearrangement. This implies that the ground $bc\bar{b}\bar{c}$ tetraquark states should be broad.
\end{abstract}

\date{\today}

\maketitle
%\tableofcontents

\newpage

%%%%%%%%%%%%%%%%%%%%%%%%%%%%%%%%%%%%%%%%%%%%%%%%%%%%%%%%%%%%%%%%%%%%%%%%%%%%%%%%%%%%%%%
\section{Introduction}
\label{sec:intro}
Exotic hadrons have become a focus of experimental and theoretical research due to their inability to be classified as conventional $q\bar{q}$ mesons or $qqq$ baryons. These states provide a unique platform for studying the non-perturbative properties of strong interactions. Over the past two decades, more than thirty potential exotic hadrons have been experimentally observed, most of them containing two heavy quarks alongside two or three light quarks. These systems, known as $XYZ$ tetraquark and $P_c$ pentaquark states ~\cite{ParticleDataGroup:2024cfk}, exhibit both long- and short-range dynamics characteristic of quantum chromodynamics (QCD), thereby complicating theoretical investigations. Moreover, there is no consensus on the structure of these exotic hadrons within the theoretical community. In contrast, fully heavy tetraquark systems are relatively simpler because they lack light quark degrees of freedom, and their dynamics are primarily dominated by perturbative QCD single-gluon exchange in the heavy quark limit ~\cite{Hughes:2017xie}. Consequently, studying fully heavy tetraquark systems is particularly valuable for understanding $XYZ$ and $P_c$ states involving light quarks.

Full-heavy tetraquark states have recently attracted significant attention. The LHCb Collaboration observed a narrow structure around 6.9 GeV, dubbed $X(6900)$, in the $J/\psi$-pair invariant mass spectrum from $pp$ collisions at center-of-mass energies of 7, 8, and 13 TeV ~\cite{LHCb:2020bwg}. This discovery, reinforced by subsequent confirmations from the CMS ~\cite{CMS:2023owd} and ATLAS ~\cite{ATLAS:2023bft} Collaborations, represents the first experimental evidence of a fully charmed tetraquark. Furthermore, additional exotic states involving four charmed (anti-)quarks, such as $X(6400)$, $X(6600)$, and $X(7200)$, have also been reported ~\cite{CMS:2023owd,ATLAS:2023bft}. However, searches for fully bottom tetraquark states in the $\Upsilon(1S)\mu^+\mu^-$ and $\Upsilon(1S)$ pair invariant mass spectra by the LHCb and CMS Collaborations, respectively, have not yielded significant evidence of narrow resonances ~\cite{CMS:2016liw,LHCb:2018uwm}.

The exploration of fully heavy tetraquarks dates back to 1976 ~\cite{Iwasaki:1976cn,Chao:1980dv,Ader:1981db} , with renewed interest in their mass spectra and decay properties. Various models, including potential models ~\cite{Bedolla:2019zwg,Deng:2020iqw,Jin:2020jfc,Becchi:2020uvq,Lu:2020cns,Giron:2020wpx,Karliner:2020dta,Zhao:2020nwy,Gordillo:2020sgc,Faustov:2020qfm,Weng:2020jao,Zhao:2020jvl,Liu:2021rtn,Zhuang:2021pci,An:2022qpt,Wang:2022yes,Zhang:2022qtp,Yu:2022lak,Wu:2024euj}, quark models ~\cite{Wang:2019rdo,Wang:2021kfv,Ortega:2023pmr,Anwar:2023fbp}, QCD sum rules ~\cite{Wang:2020ols,Zhang:2020xtb,Cao:2020gul,Wan:2020fsk,Wang:2021mma,Wu:2022qwd,Chen:2022sbf,Agaev:2023wua,Chen:2024bpz,Tang:2024zvf,Aydin:2025lbl}, effective field theories ~\cite{Wang:2020wrp,Dong:2020nwy,Guo:2020pvt,Zhu:2020snb,Zhou:2022xpd,Kuang:2023vac}, Pomeron exchange mechanisms ~\cite{Gong:2020bmg,Gong:2022hgd}, the Bethe-Salpeter (BS) equation ~\cite{Ke:2021iyh,Li:2021ygk,Santowsky:2021bhy,Zhu:2020xni}, the color flux-tube model ~\cite{Asadi:2021ids}, and nonrelativistic QCD factorization frameworks ~\cite{Feng:2023agq,Sang:2023ncm}, have been employed to interpret these states as the compact tetraquarks, dynamical rescattering mechanisms of double-charmonium channels, $c\bar{c}$ hybrids, or light Higgs-like bosons. The most widely accepted explanation involves the formation of compact tetraquark states via gluon exchange since fully heavy tetraquark states cannot involve light mesons. However, theoretical predictions of fully heavy tetraquark states still exceed experimental observations, and interpretations of their structure remain diverse.

By employing the BS equation ~\cite{Ke:2021iyh,Li:2021ygk,Santowsky:2021bhy,Zhu:2020xni}, the authors systematically studied the spectra of fully-charmed tentraquark state in a diquark and antidiquark picture. These studies uniformly suggest that $X(6900)$ cannot be the ground state of a fully charmed tetraquark, as the masses of $S$-wave fully-charmed tetraquark states are substantially lower. It is possible that $X(6900)$ is a radially excited state. Additionally, in Refs. ~\cite{Ke:2021iyh,Li:2021ygk}, the authors found that the spectra of the $S$-wave fully-charmed tetraquark states lie above the threshold of the lowest quarkonium pair $\eta_c\eta_c$ or $J/\psi J/\psi$. Thus, the ground states are expected to be broad.

In this work, we investigate the spectra of $S$-wave fully heavy $bc\bar{b}\bar{c}$ tetraquark states within the diquark-antidiquark perspective using the BS equation approach. The diquark, composed of a pair of quarks, can exist in either the $[\bar{3}]_c$ or $[6]_c$ color group representations. The $[\bar{3}]_c$ configuration is attractive in the one-gluon exchange potential, while the $[6]_c$ configuration is repulsive. Therefore, we focus exclusively on the $[\bar{3}]_c$ diquark, which consists of two different flavor quarks with possible quantum numbers $J^P=0^+$ and $1^+$. The $S$-wave $bc\bar{b}\bar{c}$ tetraquark states exhibit spin-parity quantum numbers $J^{PC}=0^{++}$, $1^{+-}$, $1^{++}$, and $2^{++}$. Motivated by potential models, the kernel for the BS equation is assumed to include scalar confinement and one-gluon exchange terms. We solve the BS equations numerically under the covariant instantaneous approximation. To account for the finite size of the heavy diquark, form factors are introduced for the effective vertex of the heavy diquark coupling to the gluon, reflecting its internal structure.

The remainder of this paper is organized as follows. In Sec.\ref{sec:BS eq}, we establish the BS equations for the ground $bc\bar{b}\bar{c}$ tetraquark states assuming that the kernels contain scalar confinement and one-gluon exchange terms in the leading order of $1/m_Q$ expansion.
The numerical solutions of the BS equations and their dependence on the parameters in our model are presented in Sec.\ref{num res}. Finally, Sec. \ref{Sec:Summary} contains a summary and discussion.

\section{The BS equation for the $bc\bar{b}\bar{c}$ system}\label{sec:BS eq}

To investigate the spectroscopy of the $bc\bar{b}\bar{c}$ system, we first construct the total wave function by combining the spatial, flavor, color, and spin wave functions. Since we focus exclusively on the $S$-wave tetraquark state within the diquark-antidiquark framework, the spatial wave function is symmetric. The quantum numbers of the tetraquark state are determined by those of the diquark and antidiquark clusters. In this study, we consider only the attractive $[\bar{3}]_c$ color configuration for the diquark, which can have quantum numbers $J^P=0^+$ or $1^+$. Consequently, all possible configurations for the $bc\bar{b}\bar{c}$ tetraquark state are summarized in Table \ref{total wave function}. We utilize the notation $\big{\vert}[bc]^{\text{color}}_{\text{spin}}[\bar{b}\bar{c}]^{\text{color}}_{\text{spin}}\big{\rangle}_{\text{spin}}$ to denote the possible wave functions.

\begin{table}[h]
\renewcommand{\arraystretch}{1.3}
\centering
\caption{Configurations of the $bc\bar{b}\bar{c}$ tetraquarks. The $S$ and $A$ represent scalar and axial-vector diquarks (or antidiquarks), respectively.}
\begin{tabular*}{\textwidth}{@{\extracolsep{\fill}}lccc}
\hline
\hline 
           & $\big{\vert}SS\big{\rangle}$ & $\big{\vert}AS\big{\rangle}$ &  $\big{\vert}AA\big{\rangle}$  \\
\hline
$0^{++}$   & $\big{\vert}[bc]^{\bar{3}}_0[\bar{b}\bar{c}]^3_0\big{\rangle}_0$      &    ——       &   $\big{\vert}[bc]^{\bar{3}}_1[\bar{b}\bar{c}]^3_1\big{\rangle}_0$      \\
$1^{+-}$   &  —— & $\frac{1}{\sqrt{2}}\left(\big{\vert}[bc]^{\bar{3}}_0[\bar{b}\bar{c}]^3_1\big{\rangle}_1-\big{\vert}[bc]^{\bar{3}}_1[\bar{b}\bar{c}]^3_0\big{\rangle}_1\right)$  &  $\vert[bc]^{\bar{3}}_1[\bar{b}\bar{c}]^3_1\rangle_1$  \\
$1^{++}$   &  ——  & $\frac{1}{\sqrt{2}}\left(\big{\vert}[bc]^{\bar{3}}_0[\bar{b}\bar{c}]^3_1\big{\rangle}_1+\big{\vert}[bc]^{\bar{3}}_1[\bar{b}\bar{c}]^3_0\big{\rangle}_1\right)$ &  —— \\ 
$2^{++}$   &  ——  & —— & $\vert[bc]^{\bar{3}}_1[\bar{b}\bar{c}]^3_1\rangle_2$ \\
\hline\hline
\end{tabular*}\label{total wave function}
\end{table}
We will subsequently establish the BS equation for each possible ground state of the $bc\bar{b}\bar{c}$ tetraquark states within the diquark-antidiquark framework.

\subsection{The tetraquark composed of a scalar diquark and a scalar antidiquark}

In a tetraquark system comprising a scalar diquark and a scalar antidiquark, a state characterized by the quantum numbers $J^{PC}=0^{++}$ can be formed. The BS wave function for this state is expressed as:
\begin{equation}
    \begin{split}
        \chi_P(x_1,x_2)=&\langle0|T\phi(x_1)\bar{\phi}(x_2)|P\rangle\\
        =&e^{-iPX}\int\frac{d^4p}{(2\pi)^4}e^{-ipx}\chi_P(p),
    \end{split}
\end{equation}
where $\phi(x_1)$ and $\bar{\phi}(x_2)$ denote the field operators for the diquark and antidiquark, respectively. The variables $X=\lambda_1x_1+\lambda_2x_2$ and $x=x_1-x_2$ represent the center-of-mass and relative coordinates of the tetraquark, respectively. The coefficients $\lambda_{1(2)}={m_{1(2)}}/({m_1+m_2})$ are determined by the masses of the diquark ($m_1$) and antidiquark ($m_2$). The momentum of the tetraquark is $P =Mv$, where $M$ is the mass of the tetraquark state and $v$ is its velocity. The variable $p$ describes the relative momentum, and consequently, the momenta of the diquark and antidiquark are given by $p_1=\lambda_1P+p$ and $p_2=\lambda_2P-p$.

Considering Lorentz covariance and parity transformation, the general form of the BS wave function for a $J^{PC}=0^{++}$ state is:
\begin{equation}\label{SS-WF}
    \chi_P(p)=s(p),
\end{equation}
where $s(p)$ is a scalar function that depends on $p^2$, $P^2$ and $(p\cdot P)^2$.
It can be demonstrated that $\chi_P(p)$ satisfies the BS equation:
\begin{equation}
    \chi_P(p)=S(p_1)\int\frac{d^4q}{(2\pi)^4}G(P,p,q)\chi_P(q)S(p_2),
\end{equation}
where $G(P,p,q)$ represents the interaction kernel, incorporating the sum of all two-particle irreducible diagrams. Using the variables $p_l=p\cdot v$ and $p_t^\mu=p^\mu-p_l v^\mu$, which are the longitudinal and transverse projections of the relative momentum along the tetraquark momentum, respectively, the scalar diquark and antidiquark propagators $S(p_1)$ and $S(p_2)$, in the leading order of $1/m_Q$ expansion, are expressed as:
\begin{equation}\label{SP1}
    S(p_1)%=\frac{i}{(\lambda_1M+p_l)^2-w_1^2+i\epsilon}
    =\frac{i}{2w_1(p_l+\lambda_1M-w_1+i\epsilon)},
\end{equation}
\begin{equation}\label{SP2}
    S(p_2)%=\frac{i}{(\lambda_2M-p_l)^2-w_2^2+i\epsilon}
    =\frac{-i}{2w_2(p_l-\lambda_2M+w_2-i\epsilon)},
\end{equation}
where $w_{1(2)}=\sqrt{m_{1(2)}^2-p_t^2}$.

Inspired by the success of potential models, scalar confinement and one-gluon exchange terms are employed in the kernel of the BS equation to analyze meson, baryon and tetraquark states~\cite{Jin:1992mw,Dai:1993np,Guo:1994fn,Guo:1998ef,Guo:1999ss,Guo:2007qu,Weng:2010rb,Yu:2018com,Feng:2013kea}. This study adopts a similar form for the kernel:
\begin{equation}\label{SS-kernel}
-iG(P,p,q)=I\otimes I V_1-\Gamma_{\mu}\otimes\Gamma^\mu V_2,
\end{equation}
where the two terms on the right hand side of Eq.(\ref{SS-kernel}) represent scalar confinement and one-gluon exchange, respectively.

\subsection{The tetraquark composed of an axial-vector diquark and a scalar antidiquark}

Here we consider a tetraquark composed of an axial-vector diquark and a scalar antidiquark, which can form states with quantum numbers $J^{PC}=1^{+-}$ and $1^{++}$. The BS wave function for such a tetraquark is expressed as:
\begin{equation}
\begin{split}
    \chi_P^{\mu}(x_1,x_2)=&\langle0|TA^\mu(x_1)\bar{\phi}(x_2)|P\rangle\\
       =&e^{-iPX}\int\frac{d^4p}{(2\pi)^4}e^{-ipx}\chi_{P}^{\mu}(p),
\end{split}
\end{equation}
where $A^\mu$ denotes the axial-vector diquark field, with its polarization vector denoted by $\epsilon_\beta$. By considering the general form of the BS wave function and its behavior under parity transformations, the BS wave functions for the $J^{PC}=1^{+-}$ and $J^{PC}=1^{++}$ tetraquark systems can be parameterized as follows:
\begin{equation}
  \chi_P^\mu(p)=a(p)\epsilon^{\mu\nu\alpha\beta}P_\nu p_\alpha\epsilon_\beta,
\end{equation}
and 
\begin{equation}
  \chi_P^\mu(p)=b(p)\epsilon^{\mu\nu\alpha\beta}P_\nu p_\alpha P\cdot p \epsilon_\beta,
\end{equation}
respectively, where the functions $a(p)$ and $b(p)$ are scalar functions. The BS equation for the tetraquark state in the momentum space can be written as:
\begin{equation}\label{BS-SV}
  \chi_P^{\mu}(p)=S^{\mu\nu}(p_1)\int\frac{d^4q}{(2\pi)^4}G_{\nu\alpha}(P,p,q)\chi_P^{\alpha}(q)S(p_2),
\end{equation}
where $S^{\mu\nu}(p_1)$ is the propagator of the axial-vector diquark, expressed in the leading order of a $1/m_Q$ expansion as:
\begin{equation}\label{VP1}
    S^{\mu\nu}(p_1)
    =-i\frac{g^{\mu\nu}-{p_1^\mu p_1^\nu}/{m_1^2}}{2w_1(p_l+\lambda_1M-w_1+i\epsilon)}.
\end{equation}
The kernel $G_{\nu\alpha}(P,p,q)$ for the BS equation involving an axial-vector diquark and a scalar antidiquark is given by
\begin{equation}\label{SV-kernel}
    iG_{\nu\alpha}=g_{\nu\alpha}I\otimes I V_1-\Gamma_{\alpha\nu\beta}\otimes\Gamma^\beta V_2,
\end{equation}
where $\Gamma_{\nu\alpha\beta}$ represents the vertex associated with a gluon interacting with two axial-vector diquarks.

\subsection{The tetraquark composed of an axial-vector diquark and an axial-vector antidiquark}

The BS wave function for a tetraquark composed of an axial-vector diquark and an axial-vector antidiquark is defined as:
\begin{equation}
\begin{split}
    \chi_P^{\mu\nu}(x_1,x_2)=&\langle0|TA^\mu(x_1)\bar{A}^\nu(x_2)|P\rangle\\
        =&e^{-iPX}\int\frac{d^4p}{(2\pi)^4}e^{-ipx}\chi_{P}^{\mu\nu}(p).
\end{split}
\end{equation}
Based on Lorentz covariance and parity constraints, the BS wave functions $\chi_P^{\mu\nu}(p)$ for the $J^{PC}=0^{++}, 1^{+-}$, and $2^{++}$ tetraquark states can generally be expanded as
\begin{equation}
    \chi_P^{\mu\nu}(p)=c_1(p)g^{\mu\nu}+c_2(p)P^\mu P^\nu+c_3(p)p^\mu p^\nu, 
\end{equation}
\begin{equation}\label{VVV-BSWF}
    \chi_P^{\mu\nu}(p)=d(p)\epsilon^{\mu\nu\rho\sigma}p_{\rho}\epsilon_{\sigma},
\end{equation}
and
\begin{equation}
\chi_P^{\mu\nu}(p)=e_1(p)\xi^{\mu\nu}+e_2(p)\xi^{\mu\sigma}p_\sigma p^\nu+e_3(p)\xi^{\nu\sigma}p_\sigma p^\mu+e_4(p)\xi^{\rho\sigma}p_\rho p_\sigma g^{\mu\nu}
+e_5(p)\xi^{\rho\sigma}p_\rho p_\sigma p^\mu p^\nu+e_6(p)\xi^{\rho\sigma}p_\rho p_\sigma P^\mu P^\nu,
\end{equation}
respectively. $c_i(p)(i=1,2,3)$, $d(p)$, and $e_j(p)(j=1,\cdots,6)$ are scalar functions. The symbols $\epsilon_{\sigma}$ and $\xi^{\mu\nu}$ represent the polarization vector and polarization tensor, respectively, satisfying the following conditions:
\begin{equation}
    \begin{split}
        P_\mu\epsilon^\mu&=0,\quad T^{\mu\nu}\equiv
        \sum_\epsilon\epsilon^\mu\epsilon^\nu=\frac{P^\mu P^\nu}{M^2}-g^{\mu\nu},\\
        \xi^{\mu\nu}&=\xi^{\nu\mu},\quad
        \xi^{\mu\nu}g_{\mu\nu}=0,\quad
        P_\mu\xi^{\mu\nu}=0,\quad
        \sum_\xi\xi^{\mu\nu}\xi^{\alpha\beta}=\frac{1}{2}\left(T^{\mu\alpha}T^{\nu\beta}+T^{\mu\beta}T^{\nu\alpha}\right)-\frac{1}{3}T^{\mu\nu}T^{\alpha\beta}.
    \end{split}
\end{equation}
Considering the constraints above and noting $p^\mu=p_t^\mu+p_lv^\mu$, it is convenient to define
\begin{equation}
    \begin{split}
        f_1&=c_1,\quad f_2=c_2+\frac{p_l^2}{M^2}P^\mu P^\nu c_3,\quad f_3=c_3,\\
        g_1&=e_1,\quad g_2=e_2,\quad g_3=e_3,\quad g_4=e_4,\quad g_5=e_5,\quad g_6=\frac{p_l^2}{M^2}P^\mu P^\nu e_5+e_6.
    \end{split}
\end{equation}
These lead to the BS wave function expressions for the $J^{PC}=0^{++}$ and $2^{++}$ states as
\begin{equation}\label{VVS-BSWF}
\chi_P^{\mu\nu}(p)=f_1(p)g^{\mu\nu}+f_2(p)P^\mu P^\nu+f_3(p)p_t^\mu p_t^\nu,
\end{equation}
and 
\begin{equation}\label{VVT-BSWF}
\chi_P^{\mu\nu}(p)=g_1(p)\xi^{\mu\nu}+g_2(p)\xi^{\mu\sigma}p_{t\sigma} p_t^\nu+g_3(p)\xi^{\nu\sigma}p_{t\sigma} p_t^\mu+g_4(p)\xi^{\rho\sigma}p_{t\rho} p_{t\sigma} g^{\mu\nu}
+g_5(p)\xi^{\rho\sigma}p_{t\rho} p_{t\sigma} p_t^\mu p_t^\nu+g_6(p)\xi^{\rho\sigma}p_{t\rho} p_{t\sigma} P^\mu P^\nu,
\end{equation}
respectively. $f_i(i=1,2,3)$ and $g_j(j=1, \cdots,6)$ are scalar functions of $p_t^2$ and $p_l$.

The general BS equation for a system composed of two axial-vector diquarks is given by
\begin{equation}\label{BS-VV}
    \chi_P^{\mu\nu}(p)=S^{\mu\alpha}(p_1)\int\frac{d^4q}{(2\pi)^4}G_{\alpha\beta\kappa\lambda}(P,p,q)\chi_P^{\kappa\lambda}(q)S^{\nu\beta}(p_2),
\end{equation}
where the propagator for heavy axial-vector antidiquark $S(p_2)$ in the $1/m_Q$ limit is expressed as:
\begin{equation}\label{VP2}
    S^{\nu\beta}(p_2)=i\frac{g^{\nu\beta}-p_2^\nu p_2^\beta/m_2^2}{2w_2(p_l-\lambda_2M+w_2-i\epsilon)},
\end{equation}
and the kernel $G_{\alpha\beta\kappa\lambda}$ for the BS equation involving double axial-vector diquarks is specified by:
\begin{equation}\label{VV-kernel}
-iG_{\alpha\beta\kappa\lambda}=g_{\alpha\kappa}g_{\beta\lambda}I\otimes I V_1-\Gamma_{\alpha\kappa\gamma}\otimes\Gamma_{\beta\lambda}^\gamma V_2.
\end{equation}

\subsection{The kernel for the BS equation}
\label{kernel}
\begin{figure}[ht]
\centering
  %\begin{center}
    \rotatebox{0}{\includegraphics*[width=0.35\textwidth]{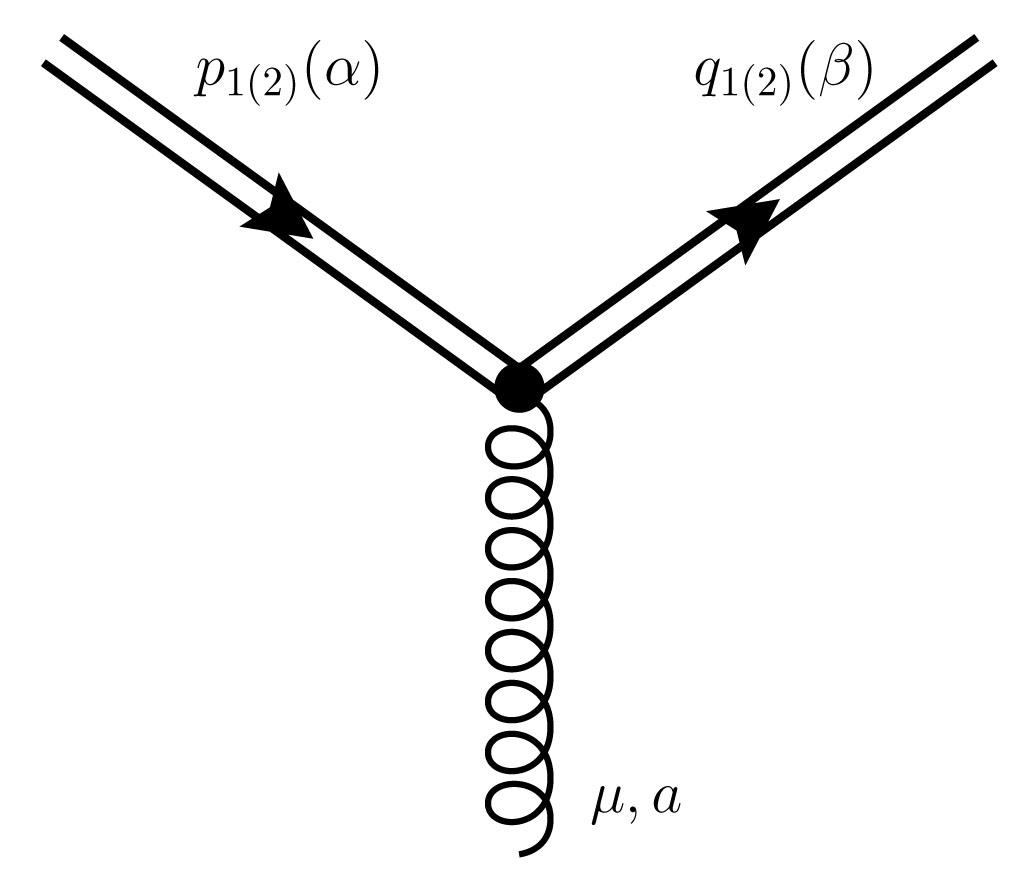}}
    \caption{The diquark-gluon-diquark vertex.}
  %\end{center}
  \label{vertex}
\end{figure}
To describe the interaction involving the non-point-like diquarks and gluon (as illustrated in Fig. \ref{vertex}), we introduce several form factors for the effective vertices to account for the diquarks' internal structure. The effective current for the scalar heavy diquark coupling to a gluon is expressed as follows  ~\cite{Anselmino:1987vk}:
\begin{equation}
    J^\mu=ig_s\frac{\lambda^a}{2}(p_1^\mu+q_1^\mu)F_s(Q^2),
\end{equation}
where $g_s$ is the strong interaction coupling constant, $F_s(Q^2)$ characterizes the internal structure of the scalar diquark, and $Q^2$ is the square of the momentum
transfer. Consequently, $\Gamma^\mu= (p_{1(2)}+q_{1(2)})^\mu F_s(Q^2)$ in Eq. (\ref{SS-kernel}).

For the axial-vector heavy diquark coupling to a gluon, the effective current is  ~\cite{Anselmino:1987vk}:
\begin{equation}
    J^{\alpha\nu\beta}=ig_s\frac{\lambda_a}{2}\left[\left(p_1+q_1\right)_\nu g_{\alpha\beta}F_{v1}(Q^2)-\left(p_{1\beta} g_{\alpha\nu}+q_{1\nu}g_{\alpha\beta}\right)F_{v2}(Q^2)+p_{1\alpha}q_{1\beta}\left(p_1+q_1\right)^\nu F_{v3}(Q^2)\right].
\end{equation}
The high momentum powers multiplied by $F_{v3}(Q^2)$ suppress its contribution at small and intermediate $Q^2$. Thus, $F_{v3}(Q^2)$ is neglected in this calculation, and $F_{v2}(Q^2)=0$ in the leading order of $1/m_Q$  expansion ~\cite{Weng:2010rb}. Hence, the effective vertex for the axial-vector heavy diquark coupling to the gluon is $\Gamma^{\alpha\nu\beta}=(p_{1(2)}+q_{1(2)})^\nu g^{\alpha\beta}F_{v1}(Q^2)$. As shown in Ref. ~\cite{Weng:2010rb}, the form factors for both the scalar and axial-vector heavy diquark vertices are equivalent in the leading order of $1/m_Q$ expansion. Therefore, we set $F_{s}(Q^2)=F_{v1}(Q^2)\equiv F(Q^2)$ for simplicity. 

From the aforementioned considerations, the diquark-gluon vertex is proportional to $(p_{1(2)}+q_{1(2)})^\mu$ and a form factor $F(Q^2)$. Following Refs. ~\cite{Guo:1998ef,Yu:2018com}, we use the form:
\begin{equation}
    F(Q^2)=\frac{\alpha_sQ_0^2}{Q^2+Q_0^2},
\end{equation}
where $Q_0$ is a parameter that ensures $F_S(Q^2)$ remains finite as $Q^2\rightarrow 0$. In the high energy region, the form factor behaves as $1/Q^2$, consistent with perturbative QCD calculations ~\cite{Lepage:1980fj}.

In the meson case, the kernel $V_1$ and $V_2$ are defined under the covariant instantaneous approximation ~\cite{Jin:1992mw,Dai:1993np,Guo:1994fn}:
\begin{equation}\label{meson-scalar-confinement}
    V_1=\frac{8\pi\kappa'}{\left[(p_t-q_t)^2+\mu^2\right]^2}-(2\pi)^3\delta^3(p_t-q_t)\int\frac{d^3k}{(2\pi)^3}\frac{8\pi\kappa}{\left[k^2+\mu^2\right]^2},
\end{equation}
and
\begin{equation}\label{meson-one-gluon-exchange}
    V_2=-\frac{16\pi}{3}\frac{\alpha_{s}}{(p_t-q_t)^2+\mu^2},
\end{equation}
where $\kappa'$ and $\alpha_s$ are the parameters associated with scalar confinement and one-gluon exchange, respectively. The counterterm in $V_1$ cancels the infrared divergence from linear confinement. It is known from the potential model ~\cite{Eichten:1978tg} and the BS equation analysis in the meson case ~\cite{Jin:1992mw,Dai:1993np,Guo:1994fn} that $\kappa'$ is approximately 0.2. The parameter $\mu$ is introduced to avoid infrared divergence in numerical calculations, with the limit $\mu\rightarrow0$ applied subsequently. 

In the tetraquark context, since confinement is still governed by scalar interaction, the form of $V_1$ remains unchanged, except for substituting $\kappa'$ with $\kappa$, which describes confinement between the diquark and antidiquark. While $\kappa'$ has dimension two, $\kappa$ has dimension four, which is due to non-perturbative diagrams that include the frozen form factor $F(Q^2)$ at low momentum. Considering the heavy masses of the diquarks and the antidiquark, we anticipate 
\begin{equation}
    \kappa\sim m_1m_2 \beta \kappa',
\end{equation}
where $\beta$ is order of one and we let it vary over a broad range from 0.5 to 2.0.

\section{Numerical results}
\label{num res}
In this section, we present the numerical solutions of the integral BS equations derived in the preceding sections. To solve the BS equations, we first substitute the propagators and interaction kernel into the BS equations. By choosing an appropriate contour, we perform the integration over the longitudinal momentum $p_l$. Subsequently, the azimuthal integration is carried out, reducing the original four-dimensional BS equation to a one-dimensional integral equation of the form
\begin{equation}
    \chi_P(|p_t|)=\int d|q_t|V(|p_t|,|q_t|)\chi_P(|q_t|).
\end{equation}
After some algebra, we find that the BS scalar wave functions $a(|p_t|)$ and $b(|b_t|)$ satisfy the same integral equation.
Then, the integration region in each integral is discretized into $n$ pieces, with $n$ being sufficiently large.  In this way, the integral equation is converted into an $n\times n$-matrix equation, and the scalar wave functions of each equation is regarded as an $n$-dimensional vector. It can be shown that the integral equations, for instance, the BS equation (\ref{BS-VV}) for the $J^{PC}=2^{++}$ state, can be illustrated as:
\begin{equation}
    \left(\begin{array}{c}
\widetilde{g}_{1}(|p_t|) \\
\widetilde{g}_{2}(|p_t|) \\
\widetilde{g}_{3}(|p_t|) \\
\widetilde{g}_{4}(|p_t|) \\
\widetilde{g}_{5}(|p_t|) \\
\widetilde{g}_{6}(|p_t|)
\end{array}\right)=\left(\begin{array}{llllll}
g_{11} & g_{12} & g_{13} & g_{14} & g_{15} & g_{16} \\
g_{21} & g_{22} & g_{23} & g_{24} & g_{25} & g_{26} \\
g_{31} & g_{32} & g_{33} & g_{34} & g_{35} & g_{36} \\
g_{41} & g_{42} & g_{43} & g_{44} & g_{45} & g_{46} \\
g_{51} & g_{52} & g_{53} & g_{54} & g_{55} & g_{56} \\
g_{61} & g_{62} & g_{63} & g_{64} & g_{65} & g_{66}
\end{array}\right)\left(\begin{array}{l}
\widetilde{g}_{1}(|q_t|) \\
\widetilde{g}_{2}(|q_t|) \\
\widetilde{g}_{3}(|q_t|) \\
\widetilde{g}_{4}(|q_t|) \\
\widetilde{g}_{5}(|q_t|) \\
\widetilde{g}_{6}(|q_t|)
\end{array}\right),
\end{equation}
where $\widetilde{g}_{i}(i=1,\cdots,6)$ is an $n$-dimensional vector, and $g_{ij}(i,j=1,\cdots,6)$ is an $n\times n$ matrix function of $|p_t|$ and $|q_t|$. Then one can obtain the numerical results of $\widetilde{g}_{i}(i=1,\cdots,6)$ by requiring the eigenvalue of the eigenvalue equation to be 1. Similar methods can be applied to evaluate other integral equations in our work.

In our model, we have several parameters, $\alpha_s$, $\beta$, $Q_0^2$, $m_{bc}$ and $E_b$. The parameter $Q_0^2$ is taken as $Q_0^2=3.2$ $\text{GeV}^2$ as in Refs. ~\cite{Anselmino:1987vk,Yu:2018com}. The values of the diquark masses (scalar diquark mass $m_{bc}^S$= 6.70 GeV and axial-vector diquark mass $m_{bc}^A$= 6.75 GeV) are used as in Ref. ~\cite{Yu:2018com}, which were obtained by fitting the experimental data of the masses of the doubly heavy baryons ~\cite{ParticleDataGroup:2024cfk}. The binding energy satisfies the relation $M_{bc\bar{b}\bar{c}}=m_{bc}+m_{\bar{b}\bar{c}}+E_b$. In general, the confinement term is still due to scalar interaction as in the meson case, the confinement parameter $\kappa'$ is taken to be 0.2 ~\cite{Jin:1992mw,Dai:1993np,Guo:1994fn}.  The value of the effective coupling constant $\alpha_s$ for fully heavy tetraquark systems in the diquark and antidiquark picture derived from QCD's running coupling behavior is not a universal constant, but depends on the specific model and energy scale. Therefore, we cannot give the definite masses of the fully-heavy tetraquark states. To explore possible solutions for the $bc\bar{b}\bar{c}$ tetraquark states, we vary $\alpha_s$ over a broader range [0.2, 0.8]. When we solve the eigenvalue euqation, the requirement that eigenvalue be 1 provides a relation between $\alpha_s$ and $\beta$. Then we obtain the parameter $\alpha_s$ for different values of $\beta$ which are listed in Table \ref{alpha-kappa}. It can be seen from the table that $\alpha$ is insensetive to $\beta$. This reflects that in the fully heavy tetraquark systems, the contribution of the scalar confinement potential is smaller than that of the one-gluon exchange potential. This is because the larger diquark and antidiquark masses reduce radius size of a fully-heavy tetraquark (like a heavy quarkonium), which pushes the system more into the regime where perturbative QCD aspects are more important than the linear confinement potential.

\begin{table}[ht]
\renewcommand{\arraystretch}{1.2}
\centering
\caption{Values of $\alpha_s$ and $\kappa$ for the tetraquark states with different $J^{PC}$.}\label{alpha-kappa}
\begin{tabular*}{\textwidth}{@{\extracolsep{\fill}}c|ccccccc}
\hline
\hline
Configuration &  $|S\bar{S}\rangle$ & \multicolumn{2}{c}{$\frac{1}{\sqrt{2}}(|A\bar{S}\rangle\pm |S\bar{A}\rangle)$}  & \multicolumn{3}{c}{$|A\bar{A}\rangle$}  \\ 
\hline
\diagbox{$\beta$}{$\alpha_s$}{$J^{PC}$} & $0^{++}$ & $1^{+-}$ & $1^{++}$ & $0^{++}$ & $1^{+-}$  & $2^{++}$ \\
\hline
0.5 & 0.46-0.80 & 0.53-0.80 & 0.53-0.80 & 0.44-0.80 & 0.46-0.80 & 0.48-0.80  \\
1.0 & 0.51-0.80 & 0.58-0.80 & 0.58-0.80 & 0.49-0.80 & 0.51-0.80 & 0.53-0.80  \\
1.5 & 0.54-0.80 & 0.61-0.80 & 0.61-0.80 & 0.53-0.80 & 0.54-0.80 & 0.55-0.80  \\
2.0 & 0.56-0.80 & 0.64-0.80 & 0.64-0.80 & 0.57-0.80 & 0.56-0.80 & 0.57-0.80  \\
\hline
\hline
\end{tabular*}
\end{table}

\begin{figure}[htb]
\centering
\subfigure[]{
\includegraphics[width=5cm]{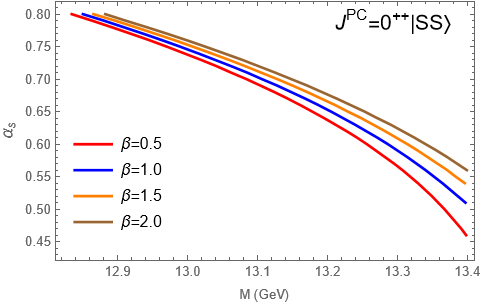}
%\caption{fig1}
}
\,
\subfigure[]{
\includegraphics[width=5cm]{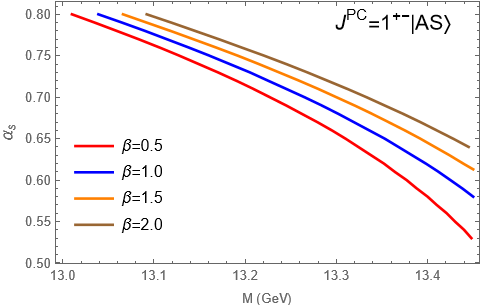}
%\caption{fig1}
}
\,
\subfigure[]{
\includegraphics[width=5cm]{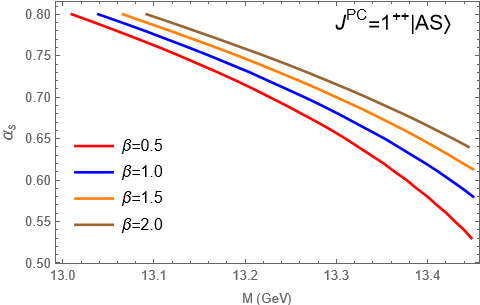}
}
\,
\subfigure[]{
\includegraphics[width=5cm]{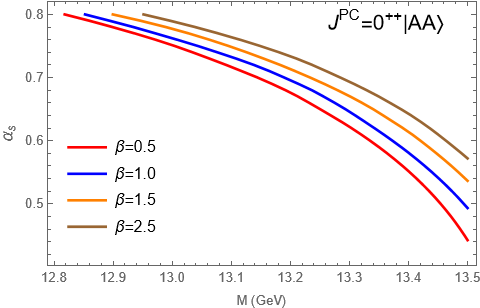}
}
\,
\subfigure[]{
\includegraphics[width=5cm]{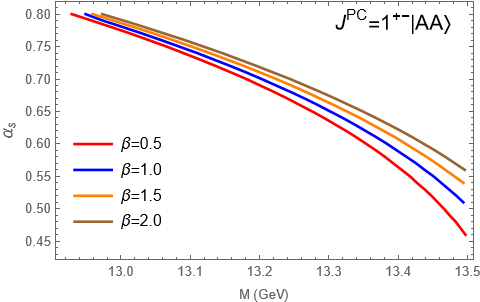}
%\caption{fig1}
}
\,
\subfigure[]{
\includegraphics[width=5cm]{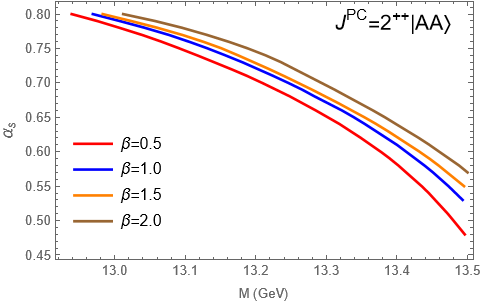}
%\caption{fig1}
}
\caption{The mass spectra of (a) $J^{PC}=0^{++} \big{\vert}SS\big{\rangle}$, (b) $J^{PC}=1^{+-} \big{\vert}AS\big{\rangle}$, (c) $J^{PC}=1^{++} \big{\vert}AS\big{\rangle}$, (d) $J^{PC}=0^{++} \big{\vert}AA\big{\rangle}$,  (e) $J^{PC}=1^{+-} \big{\vert}AA\big{\rangle}$, and (f) $J^{PC}=2^{++} \big{\vert}AA\big{\rangle}$ tetraquark states as functions of $\alpha_S$ for different $\beta$.}
\label{masses with alpha}
\end{figure}

In Fig. \ref{masses with alpha}, we display the mass spectra of $S$-wave $bc\bar{b}\bar{c}$ tetraquark states as a function of the parameter $\alpha_s$ for different $\beta$. In Table \ref{Masses}, we compare our predictions for the masses of $S$-wave $bc\bar{b}\bar{c}$ tetraquark for $\beta=1.0$ with other theoretical approaches. The values of the lowest thresholds for decays into two corresponding heavy mesons are also given in Table \ref{Masses}, the heavy meson masses are taken from PDG ~\cite{ParticleDataGroup:2024cfk}.

\begin{table}[ht]
\renewcommand{\arraystretch}{1.2}
\centering
\caption{Masses (unit: GeV) of the $S$-wave $bc\bar{b}\bar{c}$ tetraquark states with $\beta=1.0$, and the comparisons with other approaches. $E_{\text{th}}$ is the threshold of two heavy mesons.}\label{Masses}
\begin{tabular*}{\textwidth}{@{\extracolsep{\fill}}c|ccccccc}
\hline
\hline
{Configuration} &  $|S\bar{S}\rangle$ & \multicolumn{2}{c}{$\frac{1}{\sqrt{2}}(|A\bar{S}\rangle\pm |S\bar{A}\rangle)$}  & \multicolumn{3}{c}{$|A\bar{A}\rangle$}  \\ 
\hline
$J^{PC}$ & $0^{++}$ & $1^{+-}$ & $1^{++}$ & $0^{++}$ & $1^{+-}$  & $2^{++}$ \\
\hline
Our results & 12.850-13.400 & 13.039-13.450 & 13.039-13.450 & 12.824-13.500 & 12.949-13.500 & 12.969-13.500\\
~\cite{Bedolla:2019zwg} & 12.521 & 12.533 & 12.533 &  12.374 & 12.491 & 12.576  \\
~\cite{Faustov:2020qfm}  & 12.824 & 12.831  &  12.831 & 12.813 & 12.826 & 12.849  \\
~\cite{Weng:2020jao}  & 12.747 & 12.744  & 12.703 & 12.682 & 12.720 & 12.755  \\
~\cite{Zhang:2022qtp}  & 12.837 & 12.886  & 12.850 & 12.790 & 12.794 & 12.896  \\
 ~\cite{Zhang:2022qtp}  & 13.035 & 12.964  & 12.938 & 12.850 & 12.835 & 12.964  \\
~\cite{Majarshin:2021hex} & 12.359 & 12.896 & 12.155 & 12.503 & 12.016 & 12.897  \\
~\cite{Yang:2021zrc}  & $12.28^{+0.15}_{-0.14}$ & $12.32^{+0.15}_{-0.13}$  &  $12.30^{+0.15}_{-0.14}$ & $12.35^{+0.14}_{-0.12}$ & $12.38^{+0.13}_{-0.12}$ & $12.30^{+0.15}_{-0.14}$  \\
~\cite{Liu:2019zuc}  & 13.050 & 13.052  &  13.056 & 13.035 & 13.047 & 13.070  \\
~\cite{Faustov:2022mvs}  & 12.856 & 12.863  & 12.863 &  12.838 &  12.855 & 12.883  \\
~\cite{Wu:2016vtq}  & 13.553 & 13.592  & 13.510 &  13.483 &  13.520 & 13.590  \\
~\cite{Berezhnoy:2011xn}  & 12.471 &  12.488  & 12.485 &  12.359 & 12.424 & 12.566  \\
~\cite{Chen:2019vrj}  & —— &  ——  & 12.804 &  12.746 & 12.776 & 12.809  \\
  ~\cite{Malekhosseini:2025hyx}  & —— &  ——  & $12.810 \pm 0.376$ &  $12.924 \pm 0.478$ & $11.982 \pm 0.421$ & $12.276 \pm 0.329$  \\
~\cite{Malekhosseini:2025hyx} & —— &  ——  & $12.947 \pm 0.353$ &  $13.316 \pm 0.498$ & $13.165 \pm 0.458$ & $12.891 \pm 0.283$  \\
Threshold & $\eta_c(1S)\eta_b(1S)$ & $\eta_c(1S)\Upsilon(1S)$ & $J/\psi(1S)\Upsilon(1S)$ & $\eta_c(1S)\eta_b(1S)$ & $\eta_c(1S)\Upsilon(1S)$ & $J/\psi(1S)\Upsilon(1S)$  \\
$E_{\text{th}}$ & 12.3828 & 12.4445 & 12.5573 & 12.3828 & 12.4445 & 12.5573  \\
\hline
\hline
\end{tabular*}
\end{table}

From Table \ref{Masses}, it can be observed that there are considerable differences in the numerical results of $bc\bar{b}\bar{c}$ tetraquark states masses obtained from various theoretical studies. Nevertheless, the majority of these results are positioned above the corresponding lowest meson-meson thresholds. This indicates that these $bc\bar{b}\bar{c}$ tetraquark states can directly decay into the corresponding lowest meson-meson states via quark rearrangement. Additionally, our results are higher than most theoretical predictions, primarily due to the larger diquark mass obtained from our previous fitting of doubly heavy baryon masses. However, we also find that tetraquark bound states could also be formed if we had used smaller masses of heavy diquarks and antiquarks.

\begin{figure}[htb]
\centering
\subfigure[]{
\includegraphics[width=5cm]{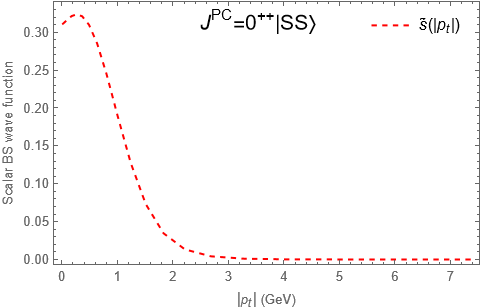}
%\caption{fig1}
}
\,
\subfigure[]{
\includegraphics[width=5cm]{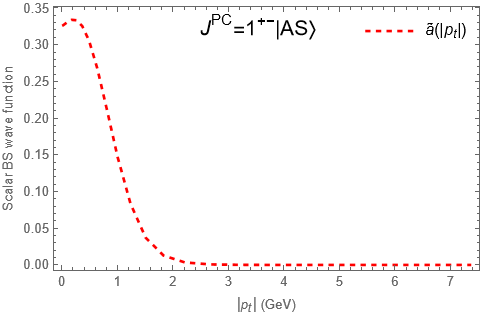}
%\caption{fig1}
}
\,
\subfigure[]{
\includegraphics[width=5cm]{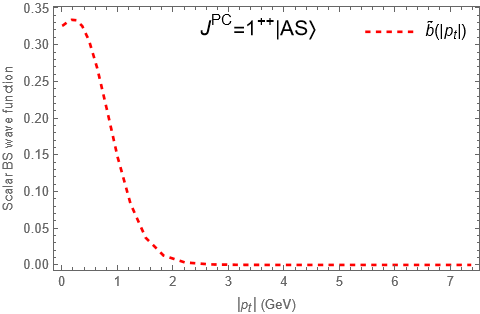}
}
\,
\subfigure[]{
\includegraphics[width=5cm]{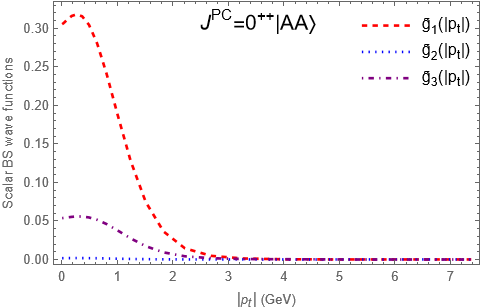}
}
\,
\subfigure[]{
\includegraphics[width=5cm]{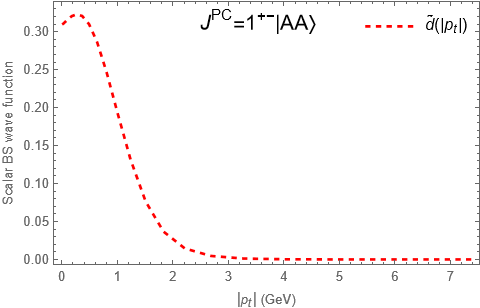}
%\caption{fig1}
}
\,
\subfigure[]{
\includegraphics[width=5cm]{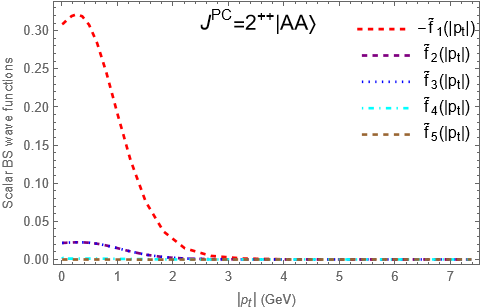}
%\caption{fig1}
}
\caption{Numerical results of the scalar wave function for (a) $J^{PC}=0^{++} \big{\vert}SS\big{\rangle}$, (b) $J^{PC}=1^{+-} \big{\vert}AS\big{\rangle}$, (c) $J^{PC}=1^{++} \big{\vert}AS\big{\rangle}$, (d) $J^{PC}=0^{++} \big{\vert}AA\big{\rangle}$,  (e) $J^{PC}=1^{+-} \big{\vert}AA\big{\rangle}$, and (f) $J^{PC}=2^{++} \big{\vert}AA\big{\rangle}$ tetraquark states with $\alpha=0.65$ and $\beta$=1.0.}
\label{wave function}
\end{figure}

The numerical results of the scalar wave functions for $S$-wave $bc\bar{b}\bar{c}$ tetraquarks with $\alpha_s$=0.65 and $\beta$=1.0 are presented in Fig. \ref{wave function}. All the scalar functions decrease to zero when $|p_t|$ is larger than about 3 GeV, because of the confinement interaction.

%%%%%%%%%%%%%%%%%%%%%%%%%%%%%%%%%   SECTION: SUMMARY  %%%%%%%%%%%%%%%%%%%%%%%%%%%%%%%%%
\section{Summary and Discussion}
\label{Sec:Summary}
In this work, we have investigated the spectra of $S$-wave $bc\bar{b}\bar{c}$ tetraquark states, which are treated as bound states of a $bc$ diquark and a $\bar{b}\bar{c}$ antidiquark. Since only the color $[\bar{3}]_c$ dconfiguration is attractive in the one-gluon exchange potential, in our analysis we have exclusively considered diquarks in the $[\bar{3}]_c$ representation. Without restrictions from the Pauli principle, the possible quantum numbers of a $[\bar{3}]_c$ diquark are $J^P=0^+$ and $J^P=1^+$. Based on this picture, we have established BS equations for the $J^{PC}=0^{++}$ $|SS\rangle$, $J^{PC}=1^{+-}$ $|AS\rangle$, $J^{PC}=1^{++}$ $|AS\rangle$, $J^{PC}=0^{++}$ $|AA\rangle$, $J^{PC}=1^{+-}$ $|AA\rangle$, and $J^{PC}=2^{++}$ $|AA\rangle$ $S$-wave $bc\bar{b}\bar{c}$ tetraquark states, in the leading order of a $1/m_Q$ expansion. The kernel of BS equations are modeled as the sum of a scalar confinement term and a one-gluon exchange term, motivated by the potential model and the BS approaches previously applied successfully to mesons, heavy baryons, and tetraquarks. Form factors have been included to account for the internal structure of the diquarks.

The BS equations have been solved numerically using the covariant instantaneous approximation. Although the BS equation is formally exact for describing bound states, its practical applications require approximations. In particular, the kernel must be modeled phenomenologically, and the diquark and antidiquark propagators are taken in their free forms, introducing certain uncertainties. Our approach involves several parameters, $\alpha_s$, $\beta$, which are varied within reasonable ranges: $\alpha_s\in[0.2,0.8]$ and $\beta\in[0.5,2.0]$, and $m_{bc}$, which are determined from fits to doubly heavy baryon mass data. 

Our numerical results indicate that $bc\bar{b}\bar{c}$ could exist. Furthermore, all of the ground states are predicted to have masses above the corresponding lowest two-meson decay thresholds, allowing strong decays via quark rearrangement. This indicates that such ground states are likely to be broad rather than narrow bound states. Additional experimental inputs from Belle-$\text{\uppercase\expandafter{\romannumeral2}}$ and LHCb will be essential to further clarify the nature of these states in the near future.

\begin{acknowledgments}
This work was supported by the National Natural Science Foundation of
China (No. 12105149, No. 12405115, No. 12475096, No. 12275024, and No. U1204115).
\end{acknowledgments}

\appendix
\section{The symmetry properties of the BS wave functions}
To obtain the symmetry properties of the BS wave functions, we need to apply these operators $P$, $C$, and $T$ to the wave functions. Here, we illustrate the procedure using the tetraquark state $|P\rangle$ with $J^{PC}=0^{++}$ composed of a scalar diquark and a scalar antidiquark as an example.  First consider parity transformation, 
\begin{equation}
    \begin{split}
        \chi_{P\zeta}(x_1,x_2)=&\langle0|T\phi(x_1)\bar{\phi}(x_2)|P\zeta\rangle\\
        =&\langle0|\mathcal{P}^{-1}\mathcal{P}T\left\{\phi(x_1)\bar{\phi}(x_2)\right\}\mathcal{P}^{-1}\mathcal{P}|P\zeta\rangle\\
        =&\eta_P\langle0|\mathcal{P}T\left\{\phi(x_1)\bar{\phi}(x_2)\right\}\mathcal{P}^{-1}|P\zeta\rangle\\
        =&\eta_P\langle0|T\left\{\phi(t_1,-\textbf{x}_1)\bar{\phi}(t_2,-\textbf{x}_2)\right\}|E,-\textbf{P},\zeta\rangle\\
        =&\eta_P\chi_{E,-\textbf{P},\zeta}(t_1,-\textbf{x}_1,t_2,-\textbf{x}_2)
    \end{split}
\end{equation}
or
\begin{equation}
   \chi_{P\zeta}(x)=\eta_P\chi_{P\zeta}(t,-\textbf{x}),
\end{equation}
where $\zeta$ is the index representing other quantum numbers of the tetraquark state.

Similarly, by applying $C$ and $T$ transformations, one can obtain the following equations:
\begin{equation}
    \chi_{P\zeta}(x_1,x_2)=\eta_C\tilde{\chi}_{P\zeta}(x_2,x_1),\quad \chi_{P\zeta}(x)=\eta_C\tilde{\chi}_{P\zeta}(-x),
\end{equation}
and 
\begin{equation}
    \chi_P(x_1,x_2)=\eta_T\bar{\chi}_P(-t_1,\textbf{x}_1,-\textbf{t}_2,x_2),\quad \chi_P(x)=\eta_T\bar{\chi}_P(-t,\textbf{x}).
\end{equation}
In momentum space, we then have
\begin{equation}
    \chi_{P\zeta}(p)=\eta_P\chi_{E,-\textbf{P},\zeta}(p_0,-\textbf{p}),
\end{equation}
\begin{equation}
    \chi_{P\zeta}(p)=\eta_C\tilde{\chi}_{P\zeta}(-p),
\end{equation}
and
\begin{equation}
    \chi_{P\zeta}(p)=\eta_T\bar{\chi}_{P\zeta}(-p_0,\textbf{p}).
\end{equation}
These equations will impose constraints on the Lorentz structure of the BS wave functions. For other states similar procedures apply.

\clearpage
\bibliography{main.bib}
\end{document}